\documentclass{article}
\usepackage{itw2005}
\usepackage{amsmath,amstext,amsthm,amssymb}
\usepackage{latex8}
\usepackage{times}

\usepackage{amsmath,amstext,amssymb,epsf}
\usepackage{epsfig}
\usepackage{verbatim}
\usepackage{pslatex}

\newcommand{\NID}{ \textsc {NID} }
\newcommand{\NGD}{ \textsc {NGD} }

\newcommand{\NCD}{ \textsc {NCD} }

\newtheorem{coro}{\sc Corollary}
\newtheorem{nota}{\sc Notation}
\newtheorem{defin}{\sc Definition}
\newtheorem{rem}{\sc Remark}
\newtheorem{cla}{\sc Claim}
\newtheorem{ex}{\sc Example}

\itwtitle{Universal Similarity}

\itwauthor{Paul Vitanyi\thanks{Part of this work was done while the author was on sabbatical leave
at National ICT of Australia, Sydney Laboratory at UNSW.
Supported in part
by the EU  EU Project RESQ IST-2001-37559,
the ESF QiT Programmme,
the EU NoE PASCAL, and the Netherlands BSIK/BRICKS project.
Address: CWI, Kruislaan 413, 1098SJ Amsterdam, The Netherlands.
{\tt paulv@cwi.nl}
}}{CWI, University of Amsterdam, National ICT of Australia}

\begin{document}
\itwmaketitle

\begin{itwabstract}
We survey a new area of parameter-free similarity distance measures
useful in data-mining,
pattern recognition, learning and automatic semantics extraction.
Given a family of distances on a set of objects,
a distance is universal up to a certain precision for that family if it
minorizes every distance in the family between every two objects 
in the set, up to the stated precision (we do not require the universal
distance to be an element of the family).
We consider similarity distances 
for two types of objects: literal objects that as such contain all of their
meaning, like genomes or books, and names for objects.
The latter may have
literal embodyments like the first type, but may also
be abstract like ``red'' or ``christianity.'' For the first type
we consider 
a family of computable distance measures
corresponding to parameters expressing similarity according to
particular features
between 
pairs of literal objects. For the second type we consider similarity
distances generated by web users corresponding to particular semantic
relations between the (names for) the designated objects. 
For both families we give universal similarity
distance measures, incorporating all particular distance measures
in the family. In the first case the universal
distance is based on compression and in the second
case it is based on Google page counts related to search terms.
In both cases experiments on a massive scale give evidence of the
viability of the approaches.
\end{itwabstract}

\begin{itwpaper}

\itwsection{Introduction}
Objects can be given literally, like the literal
four-letter genome of a mouse,
or the literal text of {\em War and Peace} by Tolstoy. For
simplicity we take it that all meaning of the object
is represented by the literal object itself. Objects can also be
given by name, like ``the four-letter genome of a mouse,''
or ``the text of {\em War and Peace} by Tolstoy.'' There are
also objects that cannot be given literally, but only by name
and acquire their meaning from their contexts in background common
knowledge in humankind, like ``home'' or ``red.''
In the literal setting, objective similarity of objects can be established
by feature analysis, one type of similarity per feature.
In the abstract ``name'' setting, all similarity must depend on 
background knowledge and common semantics relations,
which is inherently subjective and ``in the mind of the beholder.'' 

\itwsection{Compression Based Similarity}
All data are created equal but some data are more alike than others.
We have recently proposed methods expressing this alikeness,
using a new similarity metric based on compression.
It is parameter-free in that it
doesn't use any features or background knowledge about the data, and can without
changes be applied to different areas and across area boundaries.
It is universal in that it approximates the parameter
expressing similarity of the dominant feature in all pairwise
comparisons.
It is robust in the sense that its success appears independent
from the type of compressor used.
The clustering we use is hierarchical clustering in dendrograms
based on a new fast heuristic for the quartet method.
The method is available as an open-source software tool, \cite{Ci03}.

{\bf Feature-Based Similarities:}
We are presented with unknown data and
the question is to determine the similarities among them
and group like with like together. Commonly, the data are
of a certain type: music files, transaction records of ATM machines,
credit card applications, genomic data. In these data there are
hidden relations that we would like to get out in the open.
For example, from genomic data one can extract
letter- or block frequencies (the blocks are over the four-letter alphabet);
 from music files one can extract
various specific numerical features,
related to pitch, rhythm, harmony etc.
One can extract such features using for instance
Fourier transforms~\cite{TC02} or wavelet transforms~\cite{GKCwavelet},
to quantify parameters expressing similarity.
The resulting vectors corresponding to the various files are then
classified or clustered using existing classification software, based on
various standard statistical pattern recognition classifiers~\cite{TC02},
Bayesian classifiers~\cite{DTWml},
hidden Markov models~\cite{CVfolk},
ensembles of nearest-neighbor classifiers~\cite{GKCwavelet}
or neural networks~\cite{DTWml,Sneural}.
For example, in music one feature would be to look for rhythm in the sense
of beats per minute. One can make a histogram where each histogram
bin corresponds to a particular tempo in beats-per-minute and
the associated peak shows how frequent and strong that
particular periodicity was over the entire piece. In \cite{TC02}
we see a gradual change from a few high peaks to many low and spread-out
ones going from hip-hip, rock, jazz, to classical. One can use this
similarity type to try to cluster pieces in these categories.
However, such a method requires specific and detailed knowledge of
the problem area, since one needs to know what features to look for.

{\bf Non-Feature Similarities:}
Our aim
is to capture, in a single similarity metric,
{\em every effective distance\/}:
effective versions of Hamming distance, Euclidean distance,
edit distances, alignment distance, Lempel-Ziv distance,
and so on.
This metric should be so general that it works in every
domain: music, text, literature, programs, genomes, executables,
natural language determination,
equally and simultaneously.
It would be able to simultaneously detect {\em all\/}
similarities between pieces that other effective distances can detect
seperately.

Such a ``universal'' metric
was co-developed by us in \cite{LBCKKZ01,malivitch:simmet}, as a normalized
version of the ``information metric'' of \cite{liminvit:kolmbook,BGLVZ}.
Roughly speaking, two objects are deemed close if
we can significantly ``compress'' one given the information
in the other, the idea being that if two pieces are more similar,
then we can more succinctly describe one given the other.
The mathematics used is based on Kolmogorov complexity theory \cite{liminvit:kolmbook}.
In \cite{malivitch:simmet} we defined a
new class of (possibly non-metric) distances, taking values in $[0,1]$ and
appropriate for measuring effective
similarity relations between sequences, say one type of similarity
per distance, and {\em vice versa}. It was shown that an appropriately
``normalized'' information distance
minorizes every distance
in the class.
It discovers all effective similarities in the sense that if two
objects are close according to some effective similarity, then 
they are also close according to the normalized information distance.
Put differently, the normalized information distance represents
similarity according to the dominating shared feature between
the two objects being compared.
In comparisons of more than two objects, 
different pairs may have different dominating features.
For every two objects,
this universal metric distance zooms in on the dominant
similarity between those two objects
 out of a wide class of admissible similarity
features. In \cite{malivitch:simmet} we proved its optimality
and universality.
The normalized information distance also satisfies the metric 
(in)equalities, and takes values in $[0,1]$;
hence it may be called {\em ``the'' similarity metric}.

{\bf Normalized Compression Distance:}
Unfortunately, the universality of the normalized information distance
comes at the price of noncomputability, since it is based on the uncomputable
notion of Kolmogorov complexity. 
But since the Kolmogorov
complexity of a string or file is the length 
of the ultimate compressed version of that
file, 
we can use real data compression programs to approximate the Kolmogorov
complexity.
Therefore, to apply this ideal precise mathematical theory in real life,
we have to replace the use of  the noncomputable
Kolmogorov complexity by an approximation
using a standard real-world compressor. 
Thus, if $C$ is a compressor and we use $C(x)$
to denote the length of the compressed version of a string $x$,
then we arrive at the {\em Normalized Compression Distance}:
\begin{equation}\label{eq.ncd}
 \NCD(x,y) = \frac{C(xy) - \min(C(x),C(y))}{\max(C(x),C(y))},
\end{equation}
where for convenience we have replaced the pair $(x,y)$ in the formula
by the concatenation $xy$,
see \cite{malivitch:simmet,civit:cbc},
In \cite{civit:cbc} we propose axioms to capture the real-world setting,
and show that \eqref{eq.ncd}
approximates optimality.
Actually, the
\NCD is a family of compression functions parameterized 
by the given data
compressor $C$. 

{\bf Universality of NCD:} In \cite{civit:cbc} we prove that the 
\NCD is universal with respect to the family of all
admissible normalized distances---a special class that 
is argued to contain all parameters and features of
similarity that are effective.
The compression-based \NCD method to
establish a universal similarity metric \eqref{eq.ncd} among objects
given as finite binary strings 
\cite{BGLVZ,LBCKKZ01,malivitch:simmet,civit:cbc,Ke04}, and has been applied to
objects like genomes, music pieces in MIDI format, computer programs
in Ruby or C, pictures in simple bitmap formats, or time sequences such as
heart rhythm data, heterogenous data and anomaly detection.
This method is feature-free in the sense
that it doesn't analyze the files looking for particular
features; rather it analyzes all features simultaneously
and determines the similarity between every pair of objects
according to the most dominant shared feature. The crucial
point is that the method analyzes the objects themselves.
This precludes comparison of abstract notions or other objects
that don't lend themselves to direct analysis, like
emotions, colors, Socrates, Plato, Mike Bonanno and Albert Einstein.

\itwsection{Google-Based Similarity}
To make computers more intelligent one would like
to represent meaning in computer-digestable form.
Long-term and labor-intensive efforts like
the {\em Cyc} project \cite{cyc:intro} and the {\em WordNet}
project \cite{wordnet} try to establish semantic relations
between common objects, or, more precisely, {\em names} for those
objects. The idea is to create
a semantic web of such vast proportions that rudimentary intelligence
and knowledge about the real world spontaneously emerges.
This comes at the great cost of designing structures capable
of manipulating knowledge, and entering high
quality contents in these structures
by knowledgeable human experts. While the efforts are long-running
and large scale, the overall information entered is minute compared
to what is available on the world-wide-web.

The rise of the world-wide-web has enticed millions of users
to type in trillions of characters to create billions of web pages of
on average low quality contents. The sheer mass of the information
available about almost every conceivable topic makes it likely
that extremes will cancel and the majority or average is meaningful
in a low-quality approximate sense. We devise a general
method to tap the amorphous low-grade knowledge available for free
on the world-wide-web, typed in by local users aiming at personal
gratification of diverse objectives, and yet globally achieving
what is effectively the largest semantic electronic database in the world.
Moreover, this database is available for all by using any search engine
that can return aggregate page-count estimates like Google for a large
range of search-queries.

While the previous \NCD method that compares the objects themselves using
\eqref{eq.ncd} is
particularly suited to obtain knowledge about the similarity of
objects themselves, irrespective of common beliefs about such
similarities, we now develop a method that uses only the name
of an object and obtains knowledge about the similarity of objects
by tapping available information generated by multitudes of
web users.
Here we are reminded of the words of D.H. Rumsfeld \cite{Ru01}
``A trained ape can know an awful lot/
Of what is going on in this world,/
Just by punching on his mouse/
For a relatively modest cost!''
The new method is useful to extract knowledge from a given corpus of
knowledge, in this case the Google database, but not to
obtain true facts that are not common knowledge in that database.
For example, common viewpoints on the creation myths in different
religions
may be extracted by the Googling method, but contentious questions
of fact concerning the phylogeny of species can be better approached
by using the genomes of these species, rather than by opinion.

{\bf Googling for Knowledge:}
Let us start with simple intuitive justification (not to be mistaken
for a substitute of the underlying mathematics)
 of the approach we propose in \cite{CV04}.
The Google search engine indexes
around ten billion pages on the web today. Each such page can be
viewed as a set of index terms. A search for a particular index term,
say ``horse'', returns a certain number of hits (web pages where
this term occurred), say 46,700,000. The number of hits for the
search term ``rider'' is, say, 12,200,000. It is also possible to search
for the pages where both ``horse'' and ``rider'' occur. This gives,
say, 2,630,000 hits.
This can be easily put in the standard  probabilistic framework.
If $w$ is a web page and $x$ a search term, then we write $x \in w$
to mean that Google returns web page $w$ when presented with search
term $x$.
An {\em event} is a set of web pages
returned by Google after
it has been presented by a search term.
We can view the event as the collection of all contexts of
the search term, background knowledge, as induced by the
accessible web pages for the Google search engine.
If the search term is $x$, then we denote the event by ${\bf x}$,
and define ${\bf x} = \{w: x \in w \}$.
The {\em probability} $p(x)$ of an event ${\bf x }$ is
the number of web pages
in the event divided by the overall number $M$ of web pages possibly
returned by Google. Thus, $p( x)= |{\bf x}|/M$.
At the time of writing, Google searches 8,058,044,651 web pages.
Define the joint event ${\bf x}  \bigcap {\bf y} = \{ w : x,y \in w\}$
as the set of web pages returned by Google,
containing both the search term $x$ and
the search term $y$. The joint probability
$p(x,  y) = |\{ w : x,y \in w\}|/M $ is the number of
web pages in the joint event  divided by the
overall number $M$ of web pages possibly
returned by Google.
This notation also allows us to define the probability $p(x|y)$
of {\em conditional} events ${\bf x}|{\bf y}
= ({\bf x} \bigcap {\bf y})/{\bf y}$ defined by
$p(x| y) = p( x,y)/p(y)$.

In the above example we have therefore $p(horse) \approx  0.0058$,
$p(rider)$ $ \approx 0.0015$, $p(horse,rider) \approx 0.0003$.
We conclude that the probability $p(horse|rider)$
 of ``horse'' accompanying ``rider''
is $\approx 1/5$ and the probability $p(rider|horse)$ of ``rider'' accompanying
``horse'' is $\approx 1/19$.  The probabilities are asymmetric, and it is the
least probability that is the significant one. A very general search term
like ``the'' occurs in virtually all (English language) web pages.
Hence $p(the|rider) \approx 1$, and for almost all search
terms $x$ we have $p(the|x) \approx 1$. But $p(rider|the) \ll 1$,
say about equal to $p(rider)$, and gives the relevant information
about the association of the two terms.

Our first attempt therefore could be the distance
\[ D_1 (x,y) = \min \{ p(x|y),p(y|x) \}.
\]
Experimenting with this distance gives bad results. One reason
being that the differences among small probabilities have increasing
significance the smaller the probabilities involved are. Another
reason is that we deal with absolute probabilities: two notions
that have very small probabilities each and have $D_1$-distance
$\epsilon$ are much less similar than two notions that have
much larger probabilities and have the same $D_1$-distance.
To resolve the first problem we take the negative logarithm
of the items being minimized, resulting in
\[
 D_2 (x,y)  = \max \{  \log 1/p(x|y),  \log 1/p(y|x) \}.
\]
To resolve the second problem we normalize $D_2(x,y)$ by dividing
by the maximum of $\log 1/p(x), \log 1/p(y)$.
Altogether, we obtain
the following normalized distance
\[
D_3 (x,y) = \frac{ \max \{ \log 1/p(x|y),  \log 1/p(y|x) \}}
{ \max \{ \log 1/p(x) , \log 1/p(y) \}},
\]
for $p(x|y) > 0$ (and hence $p(y|x)>0$),
 and $D_3 (x,y) = \infty $ for $p(x|y)=0$ (and hence $p(y|x)=0$). Note that
$p(x|y) = p(x,y)/p(x)=0$ means that the search terms
``$x$'' and ``$y$'' never occur together.
The two conditional complexities are either both 0 or
they are both strictly positive. Moreover, if either of $p(x), p(y)$
is 0, then so are the conditional probabilities, but not necessarily
vice versa.

We note that in the conditional probabilities the total number $M$,
of web pages indexed by Google, is divided out. Therefore, the
conditional probabilities are independent of $M$, and can be
replaced by the number of pages, the {\em frequency}, returned by Google.
Define the {\em frequency} $f(x)$ of search term $x$ as the
number of pages a Google search for $x$ returns:
$f(x)= Mp(x)$, $f(x,y)=Mp(x,y)$, and $p(x|y) = f(x,y)/f(y)$.
Rewriting $D_3$ results in
our final notion, the {\em normalized
Google distance (\NGD)}, defined by
\begin{equation}\label{eq.ngd}
\NGD(x,y) = \frac{  \max \{\log f(x), \log f(y)\}  - \log f(x,y) \}}{
\log M - \min\{\log f(x), \log f(y) \}},
\end{equation}
and if $f(x),f(y)>0$ and $f(x,y)=0$ then $\NGD(x,y)= \infty$.
From \eqref{eq.ngd} we see that
\begin{enumerate}
\item
$\NGD(x,y)$ is undefined for  $f(x)=f(y)=0$;
\item
$\NGD(x,y) = \infty$ for $f(x,y)=0$ and either or both $f(x)>0$
and $f(y)>0$; and
\item
$ \NGD(x,y) \geq 0$ otherwise.
\end{enumerate}

With the Google hit numbers above, we can now compute
\[
\NGD(horse,rider)
\approx 0.443.
\]
We did the same calculation when Google indexed only one-half
of the current number of pages: 4,285,199,774. It is instructive that the
probabilities of the used search terms didn't change significantly over
this doubling of pages, with number of hits for ``horse''
equal 23,700,000, for ``rider'' equal 6,270,000, and
for ``horse, rider'' equal to 1,180,000.
 The $\NGD(horse,rider)$ we computed
in that situation was 0.460. This is in line with our contention
that the relative frequencies of web pages containing
search terms gives objective information about the semantic
relations between the search terms. If this is the case, then with
the vastness of the information accessed by Google, the
Google probabilities of search terms, and the computed \NGD's
should stabilize (be scale invariant) with a growing Google database.

The \NGD formula itself \eqref{eq.ngd} is {\em scale-invariant}. It is very important that, if
the number $M$ of pages indexed by Google grows sufficiently large,
the number of pages containing given search terms
goes to a fixed fraction of $M$, and so does the number of pages
containing conjunctions of search terms. This means that if $M$   doubles,
then so do the $f$-frequencies. For the \NGD to give us an objective
semantic relation between search terms,
it needs to become stable when the number $M$ of indexed pages grows.
Some evidence that this actually happens
is given  in the remark about the \NGD scaling properly.

\itwsection{From NCD to NGD}
{\bf The Google Distribution:}
\label{sect.google}
Let the set of singleton {\em Google search terms}
be denoted by ${\cal S}$. In the sequel we use both singleton
search terms and doubleton search terms $\{\{x,y\}: x,y \in {\cal S} \}$.
Let the set of web pages indexed (possible of being returned)
by Google be $\Omega$. The cardinality of $\Omega$ is denoted
by $M=|\Omega|$, and currently $8\cdot 10^9 \leq M \leq 9 \cdot 10^9$.
Assume that a priori all web pages are equi-probable, with the probability
of being returned by Google being $1/M$.  A subset of $\Omega$
is called an {\em event}. Every {\em  search term} $x$ usable by Google
defines a {\em singleton Google event} ${\bf x} \subseteq \Omega$ of web pages
that contain an occurrence of $x$ and are returned by Google
if we do a search for $x$.
Let $L: \Omega \rightarrow [0,1]$ be the uniform mass probability
function.
The probability of
such an event ${\bf x}$ is $L({\bf x})=|{\bf x}|/M$.
 Similarly, the {\em doubleton Google event} ${\bf x} \bigcap {\bf y}
\subseteq \Omega$ is the set of web pages returned by Google
if we do a search for pages containing both search term $x$ and
search term $y$.
The probability of this event is $L({\bf x} \bigcap {\bf y})
= |{\bf x} \bigcap {\bf y}|/M$.
We can also define the other Boolean combinations: $\neg {\bf x}=
\Omega \backslash {\bf x}$ and ${\bf x} \bigcup {\bf y} =
\Omega \backslash ( \neg {\bf x} \bigcap \neg {\bf y})$, each such event
having a probability equal to its cardinality divided by $M$.
If ${\bf e}$ is an event obtained from the basic events ${\bf x}, {\bf y},
\ldots$, corresponding to basic search terms $x,y, \ldots$,
by finitely many applications of the Boolean operations,
then the probability $L({\bf e}) = |{\bf e}|/M$.

%A {\em pseudo-probability} is a
%function $p: {\cal S}
%\rightarrow [0,1]$ such that $ 1 < \sum_{s \in {\cal S}} p(s) < \infty$.
Google events capture in a particular sense
all background knowledge about the search terms concerned available
(to Google) on the web. Therefore, it is natural
to consider code words for those events
as coding this background knowledge. However,
we cannot use the probability of the events directly to determine
a prefix code such as the Shannon-Fano code \cite{liminvit:kolmbook}.
The reason is that
the events overlap and hence the summed probability exceeds 1.
By the Kraft inequality \cite{liminvit:kolmbook} this prevents a
corresponding Shannon-Fano code.
The solution is to normalize:
We use the probability of the Google events to define a probability
mass function over the set $\{\{x,y\}: x,y \in {\cal S}\}$
of  Google search terms, both singleton and doubleton.
Define
\[
 N= \sum_{\{x,y\} \subseteq {\cal S}} |{\bf x} \bigcap
{\bf y}|,
\]
counting each singleton set and each doubleton set (by definition
unordered) once in the summation.
Since every web page that is indexed by Google contains at least
one occurrence of a search term, we have $N \geq M$. On the other hand,
web pages contain on average not more than a certain constant $\alpha$
search terms. Therefore, $N \leq \alpha M$.
Define
\begin{align}\label{eq.gpmf}
&g(x) = L({\bf x}) M/N =|{\bf x}|/N
\\&
\nonumber
g(x,y) =  L({\bf x} \bigcap {\bf y}) M/N =|{\bf x} \bigcap {\bf y}|/N.
\end{align}
Then, $\sum_{x \in {\cal S}} g(x)+ \sum_{x,y \in {\cal S}} g(x,y) = 1$.
Note that $g(x,y)$ is not a conventional joint distribution
since possibly $g(x) \neq \sum_{y \in {\cal S}} g(x,y)$.
Rather, we consider $g$ to be a probability mass
function over the sample space $\{ \{x,y\}: x,y \in {\cal S} \}$.
This $g$-distribution changes over time,
and between different samplings
from the distribution. But let us imagine that $g$ holds
in the sense of an instantaneous snapshot. The real situation
will be an approximation of this.
Given the Google machinery, these are absolute probabilities
which allow us to define the associated Shannon-Fano code for
both the singletons and the doubletons.

{\bf Normalized Google Distance}
The {\em Google code} length $G$
is defined by
\begin{align}\label{eq.gcc}
&G(x)= \log 1/g(x)
\\&
\nonumber
G(x,y)= \log 1/g(x,y) .
\end{align}
In contrast to strings $x$ where the complexity $C(x)$ represents
the length of the compressed version of $x$ using compressor $C$, for a search
term $x$ (just the name for an object rather than the object itself),
the Google code of length $G(x)$ represents the shortest expected
prefix-code word length of the associated Google event ${\bf x}$.
The expectation
is taken over the Google distribution $p$.
In this sense we can use the Google distribution as a compressor
for Google ``meaning'' associated with the search terms.
The associated \NCD, now called the
{\em normalized Google distance (\NGD)} is then defined
by \eqref{eq.ngd} with $N$ substituted for $M$, rewritten as
\begin{equation}\label{eq.NGD}
 \NGD(x,y)=\frac{G(x,y) - \min(G(x),G(y))}{\max(G(x),G(y))}.
\end{equation}
This $\NGD$ is an approximation to the $\NID$ 
using the Shannon-Fano code (Google code)
generated by the Google distribution as defining a compressor
approximating the length of the Kolmogorov code, using
the background knowledge on the web as viewed by Google
as conditional information. In experimental practice,
we consider $N$ (or $M$) as a normalization constant
that can be adjusted.  

{\bf Universality of NGD:} In the full paper \cite{CV04} we
show that \eqref{eq.ngd} and \eqref{eq.NGD} are
close in typical situations. 
Our experimental results suggest that every reasonable
(greater than any $f(x)$) value can be used for the normalizing factor  $N$,
and our
results seem  in general insensitive to this choice.  In our software, this
parameter $N$ can be adjusted as appropriate, and we often use $M$ for $N$.
In the full paper we analyze the mathematical properties of \NGD,
and  prove the universality of the Google distribution among web author based
distributions, as well as the universality of the \NGD with respect to
the family of the individual web author's \NGD's, that is, their
individual semantics relations, (with high probability)---not included here
for space reasons.

\itwsection{Applications}
\label{sect.exp}
{\bf Applications of NCD:}
We developed the CompLearn Toolkit, \cite{Ci03}, and performed
experiments in vastly different
application fields to test the quality and universality of the method.
The success of the method as reported below depends strongly on the
judicious use of encoding of the objects compared. Here one should
use common sense on what a real world compressor can do. There are
situations where our approach fails if applied in a
straightforward way.
For example: comparing text files by the same authors
in different encodings (say, Unicode and 8-bit version) is bound to fail.
For the ideal similarity metric  based on
Kolmogorov complexity as defined in \cite{malivitch:simmet}
this does not matter at all, but for
practical compressors used in the experiments it will be fatal.
Similarly, in the music experiments below we use symbolic MIDI
music file  format rather than wave format music files. The reason is that
the strings resulting from straightforward
discretizing the wave form files may be too sensitive to how we discretize.
Further research may ovecome this problem.

The \NCD is
not restricted to a specific application area, and
works across application area boundaries.
To extract a hierarchy of clusters
from the distance matrix,
we determine a dendrogram (binary tree)
by a new quartet
method and a fast heuristic to implement it.
The method is implemented and available as public software \cite{Ci03}, and is
robust under choice of different compressors.
This approach gives
the first completely automatic construction 
of the phylogeny tree based on whole mitochondrial genomes,
\cite{LBCKKZ01,malivitch:simmet}, 
a completely automatic construction of a language tree for over 50
Euro-Asian languages \cite{malivitch:simmet},
detects plagiarism in student programming assignments
\cite{SID}, gives phylogeny of chain letters \cite{BLM03}, and clusters
music \cite{cidervit:mus}.
Moreover, the method turns out to be robust under change of the underlying
compressor-types: statistical (PPMZ), Lempel-Ziv based  dictionary (gzip),
block based (bzip2), or special purpose (Gencompress).

To substantiate our claims of universality and robustness, in \cite{civit:cbc}
we report evidence of successful application in areas as diverse as
genomics, virology, languages, literature, music, handwritten digits,
astronomy, and
combinations of objects from completely different
domains, using statistical, dictionary, and block sorting compressors.
In genomics we presented new evidence for major questions
in Mammalian evolution, based on whole-mitochondrial genomic
analysis: the Eutherian orders and the Marsupionta hypothesis
against the Theria hypothesis.
Apart from the experiments reported in \cite{civit:cbc}, the clustering by
compression method reported
in this paper  has recently been used in many different areas all over
the world. One item in our group was
to analyze network traffic and cluster computer worms and virusses \cite{We04}.
Finally, recent  work \cite{Ke04} reports experiments with our method
on all time sequence data used in all the major data-mining
conferences in the last decade. Comparing the compression method
with all major methods used in those conferences they established
clear superiority of the compression method for clustering heterogenous
data, and for anomaly detection.

{\bf Applications of NGD:}
 This new method is  proposed in  \cite{CV04} to extract semantic
knowledge from the world-wide-web for both
supervised and unsupervised learning using the Google search engine
in an unconventional manner.  The approach is
novel in its unrestricted problem domain, simplicity of implementation,
and manifestly ontological underpinnings.  We give evidence of
elementary learning of the semantics of concepts, in
contrast to most prior approaches (outside of Knowledge Representation
research) that have neither the appearance nor the aim of dealing with ideas,
instead using abstract symbols that remain permanently ungrounded throughout
the machine learning application. 
The world-wide-web is the largest database on earth,
and it induces a
 probability mass function, the Google
distribution, via page counts for combinations of search queries.
This distribution allows us to tap the latent semantic knowledge
on the web.
While in the \NGD compression-based method
 one deals with the objects themselves,
in the current work we deal with just names for the objects.
In \cite{CV04}, as proof of principle, we demonstrate
positive correlations, evidencing an
underlying semantic structure, in both numerical symbol notations
and number-name words in a variety of natural languages
and contexts.
Next, we give applications in
(i) unsupervised hierarchical clustering, demonstrating the ability
to distinguish between colors and numbers, and
to distinguish between 17th century
Dutch painters;
(ii)
 supervised
concept-learning by example, using Support Vector Machines,
demonstrating the ability to understand
electrical terms, religious terms,
emergency incidents, and by conducting
 a massive experiment in understanding
WordNet categories \cite{Ci04};
and (iii) matching of meaning, in an example of 
automatic English-Spanish translation.

\begin{itwreferences}

\bibitem{BGLVZ}
C.H. Bennett, P. G\'acs, M. Li, P.M.B. Vit\'anyi, W. Zurek,
Information Distance, {\em IEEE Trans. Information Theory},
44:4(1998), 1407--1423.

\bibitem{BLM03}
C.H. Bennett, M. Li, B. Ma, Chain letters and evolutionary histories,
{\em Scientific American}, June 2003, 76--81.

\bibitem{burges:svmtut}
C.J.C. Burges.
A tutorial on support vector machines for pattern recognition,
{\em Data Mining and Knowledge Discovery}, 2:2(1998),121--167.

\bibitem{SID}
X. Chen, B. Francia, M. Li, B. McKinnon, A. Seker,
Shared information and program plagiarism detection,
{\em IEEE Trans. Inform. Th.}, 50:7(2004), 1545--1551.

\bibitem{Ci03}
R. Cilibrasi, The CompLearn Toolkit, CWI, 2003,
 http://complearn.sourceforge.net/

\bibitem{CVfolk}
W.~Chai and B.~Vercoe.
Folk music classification using hidden Markov models.
{\em Proc.~of International Conference on Artificial Intelligence}, 2001.

\bibitem{Ci04}
R. Cilibrasi, P. Vitanyi,
Automatic Meaning Discovery Using Google: 100 Experiments in Learning
WordNet Categories, 2004,
{\tt http://www.cwi.nl/$\sim$cilibrar/googlepaper/appendix.pdf}

\bibitem{cidervit:mus}
R.~Cilibrasi, R.~de~Wolf, P.~Vitanyi.
Algorithmic clustering of music based on string compression, 
{\em Computer Music J.}, 28:4(2004), 49-67.

\bibitem{civit:cbc}
R. Cilibrasi, P.M.B. Vitanyi, Clustering by compression, 
{\em IEEE Trans. Information Theory}, 51:4(2005), 1523- 1545. Also:
(preliminary version) http://www.archiv.org/abs/cs.CV/0312044

\bibitem{CV04}
R.~Cilibrasi, P.~Vitanyi,
Automatic meaning discovery using Google,
Manuscript, CWI, 2004;
http://arxiv.org/abs/cs.CL/0412098

%\bibitem{CPSV00}
%G. Cormode, M. Paterson, S. Sahinalp, and U. Vishkin.
%Communication complexity of document exchange.
%In {\em Proc. 11th ACM--SIAM Symp. on Discrete Algorithms}, 2000,
%197--206.

\bibitem{DTWml}
R.~Dannenberg, B.~Thom, and D.~Watson.
A machine learning approach to musical style recognition,    
{\em Proc.~International Computer Music Conference}, pp. 344-347, 1997.

\bibitem{google}
The basics of Google search,
 http://www.google.com/help/basics.html.

\bibitem{GKCwavelet}
M.~Grimaldi, A.~Kokaram, and P.~Cunningham.
Classifying music by genre using the wavelet packet transform
and a round-robin ensemble.
Technical report TCD-CS-2002-64, Trinity College Dublin, 2002.
http://www.cs.tcd.ie/publications/tech-reports/reports.02/TCD-CS-2002-64.pdf

%\bibitem{Kr49}
%L.G. Kraft,
%A device for quantizing, grouping and coding amplitude modulated
  %pulses.
%Master's thesis, Dept. of Electrical Engineering, M.I.T., Cambridge,
  %Mass., 1949.

\bibitem{Ke04}
E. Keogh, S. Lonardi, and C.A. Rtanamahatana, Toward parameter-free
data mining, In: {\em Proc. 10th ACM SIGKDD Intn'l Conf. Knowledge
Discovery and Data Mining}, Seattle, Washington, USA, August 22---25, 2004,
206--215.

\bibitem{Ko65}
A.N. Kolmogorov.
Three approaches to the quantitative definition of information,
{\em Problems Inform. Transmission}, 1:1(1965), 1--7.

\bibitem{Ko83}
A.N. Kolmogorov.
Combinatorial foundations of information theory and the calculus of
  probabilities,
{\em Russian Math. Surveys}, 38:4(1983), 29--40.

\bibitem{cyc:intro}
D.~B. Lenat.
Cyc: A large-scale investment in knowledge infrastructure,
{\em Comm. ACM}, 38:11(1995),33--38.

\bibitem{LBCKKZ01}
M. Li, J.H. Badger, X. Chen, S. Kwong, P. Kearney, and H. Zhang,
An information-based sequence distance and its application
to whole mitochondrial genome phylogeny,
{\em Bioinformatics}, 17:2(2001), 149--154.

\bibitem{malivitch:simmet}
M.~Li, X.~Chen, X.~Li, B.~Ma, P.~Vitanyi.
The similarity metric,
{\em IEEE Trans. Information Theory}, 50:12(2004), 3250- 3264.

\bibitem{liminvit:kolmbook}
M. Li, P. M.~B. Vitanyi.
{\em An Introduction to Kolmogorov Complexity and Its Applications},
2nd Ed.,
Springer-Verlag, New York, 1997.

\bibitem{cyc:onto}
S.~L. Reed, D.~B. Lenat.
Mapping ontologies into cyc.
{\em Proc. AAAI Conference 2002 Workshop on Ontologies for the Semantic Web},
Edmonton, Canada. http://citeseer.nj.nec.com/509238.html

\bibitem{Ru01}
D.H. Rumsfeld, The digital revolution,
originally published June 9, 2001, following a European trip.
In: H. Seely, The Poetry of D.H. Rumsfeld, 2003,
http://slate.msn.com/id/2081042/

\bibitem{Sneural}
P.~Scott.
Music classification using neural networks, 2001.\\
http://www.stanford.edu/class/ee373a/musicclassification.pdf

\bibitem{wordnet}
{G.A. Miller et.al, WordNet, 
A Lexical Database for the English Language,
Cognitive Science Lab, Princeton University.
\\http://www.cogsci.princeton.edu/$\sim$wn
}

\bibitem{TC02}
G.~Tzanetakis and P.~Cook, Music genre classification of audio signals,
{\em IEEE Transactions on Speech and Audio Processing},
10(5):293--302, 2002.

\bibitem{We04}
S. Wehner, Analyzing network traffic and worms using compression,
http://arxiv.org/abs/cs.CR/0504045

\end{itwreferences}

\end{itwpaper}
\end{document}